\documentclass[12pt]{article}
\usepackage{graphics}
\begin{document}

\title{Influence of intrinsic decoherence on nonclassical properties of the output
of a Bose-Einstein condensate}
\author{Shang-bin Li$^{1,2}$\thanks{Corresponding author. E-mail
address: sbli@zju.edu.cn} and Jing-bo Xu$^{1,2}$\\
$^{1}$Chinese Center of Advanced Science and Technology (World Labor-\\
atory), P.O.Box 8730, Beijing, People's Republic of China\\
$^{2}$Zhejiang Institute of Modern Physics and Department of Physics,\\
Zhejiang University, Hangzhou 310027, People's Republic of
China\thanks{Mailing address}}
\date{}
\maketitle

\begin{abstract}
{\normalsize We investigate nonclassical properties of the output
of a Bose-Einstein condensate in Milburn's model of intrinsic
decoherence. It is shown that the squeezing property of the atom
laser is suppressed due to decoherence. Nevertheless, if some very
special conditions were satisfied, the squeezing properties of
atom laser could be robust against the decoherence.\\


}
\end{abstract}
\newpage
\section * {I. INTRODUCTION}

Recently, the development of dilute gas Bose-Einstein condensation
(BEC) [1,2,3] has opened up the study of atom laser [4-7], which
is the matter wave analogs of optical lasers. The ideal atom laser
beam is a single frequency de Broglie wave with well-defined
intensity and phase. The first realization of a pulsed atom laser
was achieved with sodium atoms at MIT by coupling a BEC from a
magnetically trapped state to an untrapped state using a rf pulse
[8]. Then, it was repeated with long rf pulses [9] and Raman transitions [10].\\

In recent years, much attention has been focused on the problem of
nonlinear atomic optics. The four matter-wave mixing was realized
in the remarkable experiment [11] by making use of the optical
technique of Bragg diffraction to the condensate. The possibility
of optical control on the quantum statistics of the output matter
wave was pointed out in the framework of nonlinear atomic optical
optics [12]. Several dynamical analysis concerning the
nonclassical properties of the output of the trapped condensed
atoms have been presented in Refs.[13,14,15].\\

On the other hand, there has been increased interest in the
problem about decoherence of BEC and atom laser [16-19]. One fact
causing decoherence comes from the nonlinear interaction between
atoms. For a single-mode condensate the significant effect of
atomic collisions is to cause fluctuations in the energy and hence
fluctuations in the frequency, thus causing increased phase
uncertainty, eventually lead to decoherence [16]. In Ref.[19], the
influence of the decoherence on quantum coherent atomic tunneling
between two condensates is studied. It is shown that the
decoherence leads to the decay of the population difference and
the suppression of the coherent atomic tunneling. In this paper,
we would like to investigate the influence of decoherence on the
nonclassical properties of the output of a Bose-Einstein
condensate by adopting Milburn's model of intrinsic decoherence
[20]. The influence of decoherence on the nonclassical properties,
such as sub-poisson distribution and quadrature squeezing of the
atom laser beam is investigated. It is shown that, under very
special conditions, the atom laser beam may exhibit stationary
quadrature squeezing in such a decoherence model.\\

The paper is organized as follows: In Sec.II we briefly outline
the simple model describing the output coupling of the trapped
dilute condensed atoms, in which the nonlinear interaction between
the atoms and the quantized motion of atom center of mass in the
inhomogeneous magnetic field has been ignored. Then, the quantum
dynamical behavior of this system in the intrinsic decoherence
model is discussed by making use of Bogoliubov approximation. In
Sec.III, the influence of decoherence on the nonclassical
properties, such as sub-poisson distribution, quadrature squeezing
effect is investigated. In Sec.IV, there are some discussions.

\section * {II. THE MODEL}

In this section, we consider the output coupling of the trapped
dilute condensed atoms. For simplicity, the atoms are assumed to
have two states, $|T\rangle$ and $|F\rangle$, with the initial
condensation occurring in the trapped state $|T\rangle$. State
$|F\rangle$, which is typically unconfined by the magnetic trap,
is coupled to $|T\rangle$ by a one-mode squeezed optical field
tuned near resonance with the $|T\rangle\rightarrow|F\rangle$
transition. The Hamiltonian of this system is given as [14,15]
($\hbar=1$)
$$
H=\omega_0b^{\dagger}b+\omega_a{a}^{\dagger}a+\Omega(ab^{\dagger}c+a^{\dagger}bc^{\dagger}),
\eqno{(1)}
$$
where $b$($b^{\dagger}$) and $c$($c^{\dagger}$) are the
annihilation (creation) operators of bosonic atoms for the
untrapped state $|F\rangle$ and the trapped state $|T\rangle$ with
transition frequency $\omega_0$, respectively. $a$($a^{\dagger}$)
are the annihilation (creation) operators of the optical field
with frequency $\omega_a$. Here
$\Omega=\sqrt{\omega_a/2\varepsilon_{0}V}$ is the coupling
constant, and $V$ is the effective mode volume and $\varepsilon_0$
the vacuum permittivity. In the system (1), the atom-atom coupling
and the nonlinear interaction between the atoms and the
quantization motion of atomic center of mass in the trapped state
by an inhomogeneous magnetic field has been ignored. It was shown
that the above system leads to an oscillation behavior of the
quantum statistics between the optical field and
the output atomic laser beam [15].\\

In what follows, we outline the basic content of the Milburn model
of decoherence. Based on an assumption that on sufficiently short
time steps the quantum system does not evolves continuously under
unitary evolution but rather in a stochastic sequence of identical
unitary transformations, Milburn has derived the equation for the
time evolution density operator $\rho(t)$ of the quantum system
[20],
$$
\frac{d\rho(t)}{dt}=\gamma[\exp(-\frac{i}{\gamma}H)\rho(t)\exp(\frac{i}{\gamma}H)-\rho(t)],
\eqno{(2)}
$$
where $\gamma$ is the mean frequency of the unitary time step. In
the limit $\gamma\rightarrow\infty$, Eq.(2) reduces to the
ordinary von Neuman equation for the density operator. It is easy
to obtain the formal solution of Eq.(2) as follows,
$$
\rho(t)=\sum^{\infty}_{k=0}A_{k}(t)\rho(0)A^{\dagger}_{k}(t),
\eqno{(3)}
$$
where the Kraus operator $A_{k}(t)$ is given by
$$
A_{k}(t)=\frac{(\gamma{t})^{k/2}}{\sqrt{k!}}e^{-\gamma{t}/2}\exp(-i\frac{kH}{\gamma}).
\eqno{(4)}
$$
Obviously, $\sum^{\infty}_{k=0}A^{\dagger}_{k}(t)A_{k}(t)=I$.\\

We assume that the initial state of system (1) is described by
$\rho(0)=|\Psi(0)\rangle\langle\Psi(0)|$. Here,
$|\Psi(0)\rangle=|\alpha\rangle_{T}\otimes|\Phi(0)\rangle_{s}$
with $|\alpha\rangle_{T}$ the Glauber coherent state of the
operator $c$ characterizing the condensed atoms in the trapped
state $|T\rangle$;
$|\Phi(0)\rangle_s=|0\rangle_{F}\otimes|\xi\rangle$, $|0\rangle_F$
represents that there is initially no occupying atoms in the
untrapped state $|F\rangle$, and the optical field is in the
squeezed vacuum state $|\xi\rangle=S(\xi)|0\rangle$.
$S(\xi)=\exp(\xi{a}^{\dagger2}-\xi^{\ast}a^2)$ is the squeezed
operator.\\
Substituting the $\rho(0)$ into Eq.(3), we obtain
$$
\rho(t)=e^{-\gamma{t}}\exp(\alpha{c}^{\dagger}-\alpha^{\ast}c)\sum^{\infty}_{k=0}\frac{(\gamma{t})^k}{k!}M^{k}|0\rangle_{TT}\langle0|\otimes|\Phi(0)\rangle_{ss}\langle\Phi(0)|M^{\dagger{k}}\exp(-\alpha{c}^{\dagger}+\alpha^{\ast}c)
\eqno{(5)}
$$
where
$$
M^{k}=\exp[-i\frac{k}{\gamma}(H_0+H_1)],
\eqno{(6)}
$$
$$
H_0=\omega_0b^{\dagger}b+\omega_a{a}^{\dagger}a+\Omega(\alpha{a}b^{\dagger}+\alpha^{\ast}a^{\dagger}b),
\eqno{(7)}
$$
$$
H_1=\Omega(acb^{\dagger}+a^{\dagger}c^{\dagger}b).
\eqno{(8)}
$$
If $|\alpha|\gg1$, we can adopt the Bogoliubov approximation [21],
i.e., neglect $H_1$ in Eq.(5). Then in the Bogoliubov
approximation, the density operator $\rho(t)$ can be reexpressed
as follows,
$$
\rho(t)=|\alpha\rangle_{TT}\langle\alpha|\otimes\rho_s(t),
\eqno{(9)}
$$
where
$$
\rho_s(t)=e^{-\gamma{t}}\sum^{\infty}_{k=0}\frac{(\gamma{t})^k}{k!}\exp(-i\frac{kH_0}{\gamma})|0\rangle_{FF}\langle0|\otimes|\xi\rangle\langle\xi|\exp(i\frac{kH_0}{\gamma}),
\eqno{(10)}
$$
is the reduced density operator describing the subsystem of the
untrapped atoms and optical field.\\

Now, we confine ourselves in the resonance case, i.e.,
$\omega_0=\omega_a=\omega$. We define the operators $a(k)$
($a^{\dagger}(k)$) and $b(k)$ ($b^{\dagger}(k)$) as
$$
a(k)=\exp(\frac{ikH_0}{\gamma})a\exp(-\frac{ikH_0}{\gamma}),
$$
$$
a^{\dagger}(k)=\exp(\frac{ikH_0}{\gamma})a^{\dagger}\exp(-\frac{ikH_0}{\gamma}),
$$
$$
b(k)=\exp(\frac{ikH_0}{\gamma})b\exp(-\frac{ikH_0}{\gamma}),
$$
$$
b^{\dagger}(k)=\exp(\frac{ikH_0}{\gamma})b^{\dagger}\exp(-\frac{ikH_0}{\gamma}).
\eqno{(11)}
$$
It is easy to obtain that
$$
a(k)=\cos(\frac{k\Omega^{\prime}}{\gamma})e^{-ik\omega/\gamma}a-i\sin(\frac{k\Omega^{\prime}}{\gamma})e^{i(\theta-k\omega/\gamma)}b,
$$
$$
a^{\dagger}(k)=\cos(\frac{k\Omega^{\prime}}{\gamma})e^{ik\omega/\gamma}a^{\dagger}+i\sin(\frac{k\Omega^{\prime}}{\gamma})e^{-i(\theta-k\omega/\gamma)}b^{\dagger},
$$
$$
b(k)=\cos(\frac{k\Omega^{\prime}}{\gamma})e^{-ik\omega/\gamma}b-i\sin(\frac{k\Omega^{\prime}}{\gamma})e^{-i(\theta+k\omega/\gamma)}a,
$$
$$
b^{\dagger}(k)=\cos(\frac{k\Omega^{\prime}}{\gamma})e^{ik\omega/\gamma}b^{\dagger}+i\sin(\frac{k\Omega^{\prime}}{\gamma})e^{i(\theta+k\omega/\gamma)}a^{\dagger},
\eqno{(12)}
$$
where $\Omega^{\prime}=|\alpha|\Omega$ and
$e^{-i\theta}=\alpha/|\alpha|$. By making use of Eq.(12), we can
express average values of arbitrary operator functionals
$G(a,a^{\dagger};b,b^{\dagger})$ as following
$$
\textrm{Tr}[G(a,a^{\dagger};b,b^{\dagger})\rho_s(t)]=e^{-\gamma{t}}\sum^{\infty}_{k=0}\frac{(\gamma{t})^k}{k!}\textrm{Tr}[G(a(k),a^{\dagger}(k);b(k),b^{\dagger}(k))|0\rangle_{FF}\langle0|\otimes|\xi\rangle\langle\xi|].
\eqno{(13)}
$$
\section * {III. THE INFLUENCE OF INTRINSIC DECOHERENCE ON THE NONCLASSICAL PROPERTIES OF THE ATOM LASER BEAM}
In this section, we investigate the nonclassical properties of the
atom laser in the Milburn's model of decoherence. The average
numbers and the fluctuation of the output atoms as well as the
out-state photon can be obtained by making use of the Eq.(13)
$$
N_a(t)=\textrm{Tr}(a^{\dagger}a\rho_s(t))
$$
$$
~~~=[\frac{1}{2}+\frac{1}{4}\exp(\gamma{t}e^{2i\Omega^{\prime}/\gamma}-\gamma{t})+\frac{1}{4}\exp(\gamma{t}e^{-2i\Omega^{\prime}/\gamma}-\gamma{t})]\sinh^2r,
\eqno{(14)}
$$
$$
N_b(t)=\textrm{Tr}(b^{\dagger}b\rho_s(t))
$$
$$
~~~=[\frac{1}{2}-\frac{1}{4}\exp(\gamma{t}e^{2i\Omega^{\prime}/\gamma}-\gamma{t})-\frac{1}{4}\exp(\gamma{t}e^{-2i\Omega^{\prime}/\gamma}-\gamma{t})]\sinh^2r,
\eqno{(15)}
$$
$$
\triangle{N}^2_a(t)=\textrm{Tr}(a^{\dagger}aa^{\dagger}a\rho_s(t))-[\textrm{Tr}(a^{\dagger}a\rho_s(t))]^2
$$
$$
=\{\frac{3}{4}+e^{-\gamma{t}}[\frac{1}{2}\exp(\gamma{t}e^{2i\Omega^{\prime}/\gamma})+\frac{1}{2}\exp(\gamma{t}e^{-2i\Omega^{\prime}/\gamma})+\frac{1}{8}\exp(\gamma{t}e^{4i\Omega^{\prime}/\gamma})+\frac{1}{8}\exp(\gamma{t}e^{-4i\Omega^{\prime}/\gamma})]\}\sinh^2r\cosh^2r
$$
$$
+\{\frac{1}{8}+\frac{1}{16}e^{-\gamma{t}}[\exp(\gamma{t}e^{4i\Omega^{\prime}/\gamma})+\exp(\gamma{t}e^{-4i\Omega^{\prime}/\gamma})]
$$
$$
-\frac{1}{16}e^{-2\gamma{t}}[\exp(2\gamma{t}e^{2i\Omega^{\prime}/\gamma})+\exp(2\gamma{t}e^{-2i\Omega^{\prime}/\gamma})+2\exp(2\gamma{t}\cos(2\Omega^{\prime}/\gamma))]\}\sinh^4r
$$
$$
+\{\frac{1}{8}-\frac{1}{16}e^{-\gamma{t}}[\exp(\gamma{t}e^{4i\Omega^{\prime}/\gamma})+\exp(\gamma{t}e^{-4i\Omega^{\prime}/\gamma})]\}\sinh^2r,
\eqno{(16)}
$$
$$
\triangle{N}^2_b(t)=\textrm{Tr}(b^{\dagger}bb^{\dagger}b\rho_s(t))-[\textrm{Tr}(b^{\dagger}b\rho_s(t))]^2
$$
$$
=\{\frac{3}{4}-e^{-\gamma{t}}[\frac{1}{2}\exp(\gamma{t}e^{2i\Omega^{\prime}/\gamma})+\frac{1}{2}\exp(\gamma{t}e^{-2i\Omega^{\prime}/\gamma})-\frac{1}{8}\exp(\gamma{t}e^{4i\Omega^{\prime}/\gamma})-\frac{1}{8}\exp(\gamma{t}e^{-4i\Omega^{\prime}/\gamma})]\}\sinh^2r\cosh^2r
$$
$$
+\{\frac{1}{8}+\frac{1}{16}e^{-\gamma{t}}[\exp(\gamma{t}e^{4i\Omega^{\prime}/\gamma})+\exp(\gamma{t}e^{-4i\Omega^{\prime}/\gamma})]
$$
$$
-\frac{1}{16}e^{-2\gamma{t}}[\exp(2\gamma{t}e^{2i\Omega^{\prime}/\gamma})+\exp(2\gamma{t}e^{-2i\Omega^{\prime}/\gamma})+2\exp(2\gamma{t}\cos(2\Omega^{\prime}/\gamma))]\}\sinh^4r
$$
$$
+\{\frac{1}{8}-\frac{1}{16}e^{-\gamma{t}}[\exp(\gamma{t}e^{4i\Omega^{\prime}/\gamma})+\exp(\gamma{t}e^{-4i\Omega^{\prime}/\gamma})]\}\sinh^2r,
\eqno{(17)}
$$
where $r=2|\xi|$.\\

In order to discuss the quantum statistical properties of the atom
laser and the out-state optical field, we can calculate the Mandel
Q parameters defined as [22]
$$
Q_i(t)=\frac{\triangle{N}^2_i(t)-N_i(t)}{N_i(t)},~~~(i=a,b)
\eqno{(18)}
$$
$Q_i(t)<0$, $Q_i(t)=0$ or $Q_i(t)>0$ mean the quantum state of the
optical field or the atom laser field satisfy sub-Poisson, Poisson
or super-Poisson distribution, respectively. In Fig.1 and Fig.2,
the Mandel Q parameters $Q_a$ and $Q_b$ are plotted as a function
of the time $t$ for two different values of the parameter
$\gamma$, respectively. We can observe that both the optical field
and the atom laser beam satisfy the super-poisson distribution for
any $t>0$ in the cases with finite values of the parameter
$\gamma$.
\\

We now turn to discuss the influence of intrinsic decoherence on
quadrature squeezing of atom laser and the optical field. We
introduce four quadrature operators defined by
$$
X^{(a)}_1=\frac{1}{2}(a+a^{\dagger}),~~~X^{(a)}_2=\frac{1}{2i}(a-a^{\dagger}),~~~X^{(b)}_1=\frac{1}{2}(b+b^{\dagger}),~~~X^{(b)}_2=\frac{1}{2i}(b-b^{\dagger})
\eqno{(19)}
$$
These operators satisfy the commutation relation
$$
[X^{(a)}_1,X^{(a)}_2]=\frac{i}{2},~~~[X^{(b)}_1,X^{(b)}_2]=\frac{i}{2},
\eqno{(20)}
$$
which implies the Heisenberg uncertainly relations
$$
\langle(\triangle{X}^{(a)}_1)^2\rangle\langle(\triangle{X}^{(a)}_2)^2\rangle\geq\frac{1}{16},~~~\langle(\triangle{X}^{(b)}_1)^2\rangle\langle(\triangle{X}^{(b)}_2)^2\rangle\geq\frac{1}{16}.
\eqno{(21)}
$$
Squeezing is said to exist whenever
$\langle(\triangle{X}^{(j)}_i)^2\rangle<{1/4}$, ($i=1,2$),
($j=a,b$). In order to characterize the influence of intrinsic
decoherence on the quadrature squeezing of the atom laser and
optical field, we calculate the following squeezing coefficients
[23]
$$
S^{(j)}_i=\frac{\langle(\triangle{X}^{(j)}_i)^2\rangle-0.25}{0.25},
\eqno{(22)}
$$
where $-1\leq{S}^{(j)}_i<0$ for quadrature squeezing. We express
the squeezing coefficients as follows
$$
S^{(a)}_1=2N_a(t)+\textrm{Re}\{e^{-\gamma{t}}[\exp(\gamma{t}e^{-2i\omega/\gamma})+\frac{1}{2}\exp(\gamma{t}e^{2i(\Omega^{\prime}-\omega)/\gamma})+\frac{1}{2}\exp(\gamma{t}e^{-2i(\Omega^{\prime}+\omega)/\gamma})]\sinh{r}\cosh{r}e^{i\phi}\},
\eqno{(23)}
$$
$$
S^{(a)}_2=2N_a(t)-\textrm{Re}\{e^{-\gamma{t}}[\exp(\gamma{t}e^{-2i\omega/\gamma})+\frac{1}{2}\exp(\gamma{t}e^{2i(\Omega^{\prime}-\omega)/\gamma})+\frac{1}{2}\exp(\gamma{t}e^{-2i(\Omega^{\prime}+\omega)/\gamma})]\sinh{r}\cosh{r}e^{i\phi}\},
\eqno{(24)}
$$
$$
S^{(b)}_1=2N_b(t)+\textrm{Re}\{e^{-\gamma{t}-2i\theta}[-\exp(\gamma{t}e^{-2i\omega/\gamma})
$$
$$
~~~+\frac{1}{2}\exp(\gamma{t}e^{2i(\Omega^{\prime}-\omega)/\gamma})+\frac{1}{2}\exp(\gamma{t}e^{-2i(\Omega^{\prime}+\omega)/\gamma})]\sinh{r}\cosh{r}e^{i\phi}\},
\eqno{(25)}
$$
$$
S^{(b)}_2=2N_b(t)-\textrm{Re}\{e^{-\gamma{t}-2i\theta}[-\exp(\gamma{t}e^{-2i\omega/\gamma})
$$
$$
~~~+\frac{1}{2}\exp(\gamma{t}e^{2i(\Omega^{\prime}-\omega)/\gamma})+\frac{1}{2}\exp(\gamma{t}e^{-2i(\Omega^{\prime}+\omega)/\gamma})]\sinh{r}\cosh{r}e^{i\phi}\},
\eqno{(26)}
$$
where $e^{i\phi}=\xi/|\xi|$. It is obvious that
$S^{(b)}_1+S^{(b)}_2=4N_b(t)$, and $N_b(t)$ is always
non-negative. So, we need only investigate $S^{(b)}_1$ or
$S^{(b)}_2$ to explore the squeezing property of the atom laser.
In what follows, it is assumed $\phi=\theta=0$. We start our
analysis of the squeezing properties of both the atom laser and
the optical field in the limit case with $\gamma\rightarrow\infty$
and finite values of $\omega$ and $\Omega^{\prime}$, which means
not any decoherence is presented. Then, the Eqs.(23-26) reduces to
the results in Ref.[15], in which the squeezing coefficients of
both the atom laser and the optical field exhibit the Rabi-like
oscillation. If $\gamma$ is a finite value but remains large, i.e.
$\omega/\gamma\ll1$ and $\Omega^{\prime}/\gamma\ll1$, the
expressions of the squeezing coefficients $S^{(a)}_i$ and
$S^{(b)}_i$ ($i=1,2$) can be approximated as
$$
S^{(a)}_1\approx[1+\cos2\Omega^{\prime}t\exp(-2\Omega^{\prime2}t/\gamma)]\sinh^2r+\{\cos2\omega{t}\exp(-2\omega^{2}t/\gamma)
$$
$$
+\frac{1}{2}\cos[2(\Omega^{\prime}-\omega)t]\exp[-2(\Omega^{\prime}-\omega)^2t/\gamma]+\frac{1}{2}\cos[2(\Omega^{\prime}+\omega)t]\exp[-2(\Omega^{\prime}+\omega)^2t/\gamma]\}\sinh{r}\cosh{r},
\eqno{(27)}
$$
$$
S^{(a)}_2\approx[1+\cos2\Omega^{\prime}t\exp(-2\Omega^{\prime2}t/\gamma)]\sinh^2r-\{\cos2\omega{t}\exp(-2\omega^{2}t/\gamma)
$$
$$
+\frac{1}{2}\cos[2(\Omega^{\prime}-\omega)t]\exp[-2(\Omega^{\prime}-\omega)^2t/\gamma]+\frac{1}{2}\cos[2(\Omega^{\prime}+\omega)t]\exp[-2(\Omega^{\prime}+\omega)^2t/\gamma]\}\sinh{r}\cosh{r},
\eqno{(28)}
$$
$$
S^{(b)}_1\approx[1-\cos2\Omega^{\prime}t\exp(-2\Omega^{\prime2}t/\gamma)]\sinh^2r+\{-\cos2\omega{t}\exp(-2\omega^{2}t/\gamma)
$$
$$
+\frac{1}{2}\cos[2(\Omega^{\prime}-\omega)t]\exp[-2(\Omega^{\prime}-\omega)^2t/\gamma]+\frac{1}{2}\cos[2(\Omega^{\prime}+\omega)t]\exp[-2(\Omega^{\prime}+\omega)^2t/\gamma]\}\sinh{r}\cosh{r},
\eqno{(29)}
$$
$$
S^{(b)}_2\approx[1-\cos2\Omega^{\prime}t\exp(-2\Omega^{\prime2}t/\gamma)]\sinh^2r-\{-\cos2\omega{t}\exp(-2\omega^{2}t/\gamma)
$$
$$
+\frac{1}{2}\cos[2(\Omega^{\prime}-\omega)t]\exp[-2(\Omega^{\prime}-\omega)^2t/\gamma]+\frac{1}{2}\cos[2(\Omega^{\prime}+\omega)t]\exp[-2(\Omega^{\prime}+\omega)^2t/\gamma]\}\sinh{r}\cosh{r},
\eqno{(30)}
$$
From Eqs.(27-30), we find that all of the squeezing coefficients
$S^{(a)}_i$ and $S^{(b)}_i$ ($i=1,2$) tend to a fixed positive
value $\sinh^2r$ as the time $t\rightarrow\infty$, except that
they tend to $\sinh^2r\pm\frac{1}{2}\sinh{r}\cosh{r}$ in the
special case with $\Omega^{\prime}=\omega$. With the further
decrease of $\gamma$, the oscillatory behaviors of the squeezing
properties of both the atom laser and the optical field become
frozen. In Fig.3, the squeezing coefficient $S^{(b)}_2$ is plotted
as a function of time $t$ for three different values of parameter
$\gamma$. With the decreases of parameter $\gamma$, we can observe
rapid deterioration of the Rabi-like oscillation of squeezing
coefficient. In Fig.3(b) and Fig3(c), if the time $t$ become very
large, the squeezing coefficient $S^{(b)}_2$ tends to be a fixed
positive value $0.093$, which means the decoherence eventually
completely destroy the squeezing property of the atom laser.
However, under very special conditions that
$\Omega^{\prime}=\omega\ll\gamma$, $0<\tanh{r}<\frac{1}{2}$ and
$\gamma$ is large but remain finite, $S^{(b)}_2$ will tend to be a
negative value as the time $t$ approaches to infinite, which means
the stationary state of the
atom laser gets squeezed. This can be clearly seen from Fig.4.\\

Recently, the sensitivity of quantum systems that are chaotic in a
classical limit to small perturbations has been investigated in
Ref.[24], and the relation between the sensitive and decoherence
has been discussed. In what follows, we briefly investigate
influence of small perturbation on the quadrature squeezing of
atom laser. From Eq.(26), we can observe that the dynamical
behavior of $S^{(b)}_2$ is dependent on the values of the
exponential facts
$\gamma[{e}^{2i(\Omega^{\prime}-\omega)/\gamma}-1]$,
$\gamma[{e}^{\pm2i\Omega^{\prime}/\gamma}-1]$,
$\gamma[{e}^{-2i\omega/\gamma}-1]$, and
$\gamma[{e}^{-2i(\Omega^{\prime}+\omega)/\gamma}-1]$. In the case
with $\Omega^{\prime}\simeq\omega\ll\gamma$, the term
$\gamma[{e}^{2i(\Omega^{\prime}-\omega)/\gamma}-1]\simeq(2i\delta-2\delta^2/\gamma)$
plays a dominantly role in the long time dynamical behavior, where
$\delta=\Omega^{\prime}-\omega$. In Fig.5, we plotted $S^{(b)}_2$
as the function of time $t$ for different small values of
$\delta$. Since $\Omega^{\prime}$ is dependent on the amplitude
$\alpha$ of the initial coherent state of condensed atoms in the
trapped state $|T\rangle$, we expect that the squeezing dynamical
behavior of the atom laser is sensitive to the initial state
of trapped condensed atoms, just as shown in Fig.5. \\

\section * {IV. DISCUSSION}
\hspace*{8mm}In this paper, we investigate nonclassical properties
of the output of a Bose-Einstein condensate in Milburn's model of
intrinsic decoherence by making use of Bogoliubov approximation.
It is shown that the squeezing properties of the atom laser field
can be destroyed by the decoherence under most conditions.
However, under some very special conditions, the squeezing
properties of atom laser is robust against the decoherence. This
phenomenon is highly dependent on the particular
modelling of decoherence.\\

In what follows, we briefly discuss the relevance of our
theoretical results to the realistic experimental conditions. By
making use of the preliminary atom laser experiments [8], as a
guide to realistic parameter values, we choose $\omega=300$kHz,
$\Omega=60$kHz, the initial trapped atom number
$N=|\alpha|^2=10^6$, the mean frequency of the unitary time step
$\gamma=10^5$MHz, and $r=1$ as a example to illustrate our
theoretical results. In such a condition, the Rabi-like
oscillation behaviors of the numbers of both the output atom and
output photon become frozen after $100\mu{s}$. However, the atom
laser persists in exhibiting oscillation of squeezing coefficient
and its squeezing property is finally destroyed by decoherence
till about $0.15s$.\\

Milburn's model of intrinsic decoherence is currently a very
active field of research. Nevertheless, so far little attention
has been paid to its study in the context of ultracold quantum
gases. However, due to their high sensitivity to decoherence, such
systems could provide an interesting testing ground for the
Milburn model. In this paper, we provide a first step in this
direction by studying the influence of intrinsic decoherence on
the output properties of a simple atom laser model. A thorough
discussion of all other sources of decoherence that could be
present in an experiment and comparison with the results presented
in this work should be interesting and necessary in the future
study. It is also very interesting to investigate the entanglement
between output atoms and output photons and discuss the influence
of intrinsic decoherence on entanglement [25-28]. Moreover, it is
worth to discuss the possible quantum chaotic dynamical behavior
of atom laser and the quantum-classical corresponding by fully
taking account of the atom-atom interaction.
\section * {ACNOWLEDGMENTS}
This project was supported by the National Natural Science
Foundation of China (Project NO. 10174066).

\newpage
{\Large\bf Figure Caption}
\begin{description}
\item[FIG.1.]The Mandel Q parameter $Q_a$ of optical field is plotted as a function of time $t$ with
$\Omega^{\prime}=1$ and $r=2$ for two different values of the parameter $\gamma$,
(Solid line) $\gamma=\infty$, (Dot line) $\gamma=10^2$.\\
\item[FIG.2.]The Mandel Q parameter $Q_b$ of atom laser field is plotted as a function of time $t$ with
$\Omega^{\prime}=1$ and $r=2$ for two different values of the
parameter $\gamma$,
(Solid line) $\gamma=\infty$, (Dot line) $\gamma=10^2$.\\
\item[FIG.3.]The quadrature squeezing coefficient $S^{(b)}_2$ of the
atom laser field is plotted as a function of time $t$ with
$\omega=0.1$, $\phi=0$, $\theta=0$, $\Omega^{\prime}=\pi$ and
$r=0.3$ for three different values of the parameter $\gamma$, (a)
$\gamma=\infty$,
(b) $\gamma=10^3$, (c) $\gamma=10^2$.\\
\item[FIG.4.]The quadrature squeezing coefficient $S^{(b)}_2$ of the
atom laser field is plotted as a function of time $t$ with
$\omega=10$, $\phi=0$, $\theta=0$, $\Omega^{\prime}=10$, $r=0.3$ and $\gamma=10^2$.\\
\item[FIG.5.]The quadrature squeezing coefficient $S^{(b)}_2$ of the
atom laser field is plotted as a function of time $t$ with
$\omega=10$, $\phi=0$, $\theta=0$, $r=0.4$ and $\gamma=10^2$ for
four different values of $\Omega^{\prime}$: (Solid line)
$\Omega^{\prime}=10$, (Dot line) $\Omega^{\prime}=10+10^{-7}$,
(Dash Dot line) $\Omega^{\prime}=10+2\times10^{-7}$, (Dash Dot Dot
line) $\Omega^{\prime}=10+3\times10^{-7}$.\\
\end{description}

\begin{figure}
\centering
\includegraphics{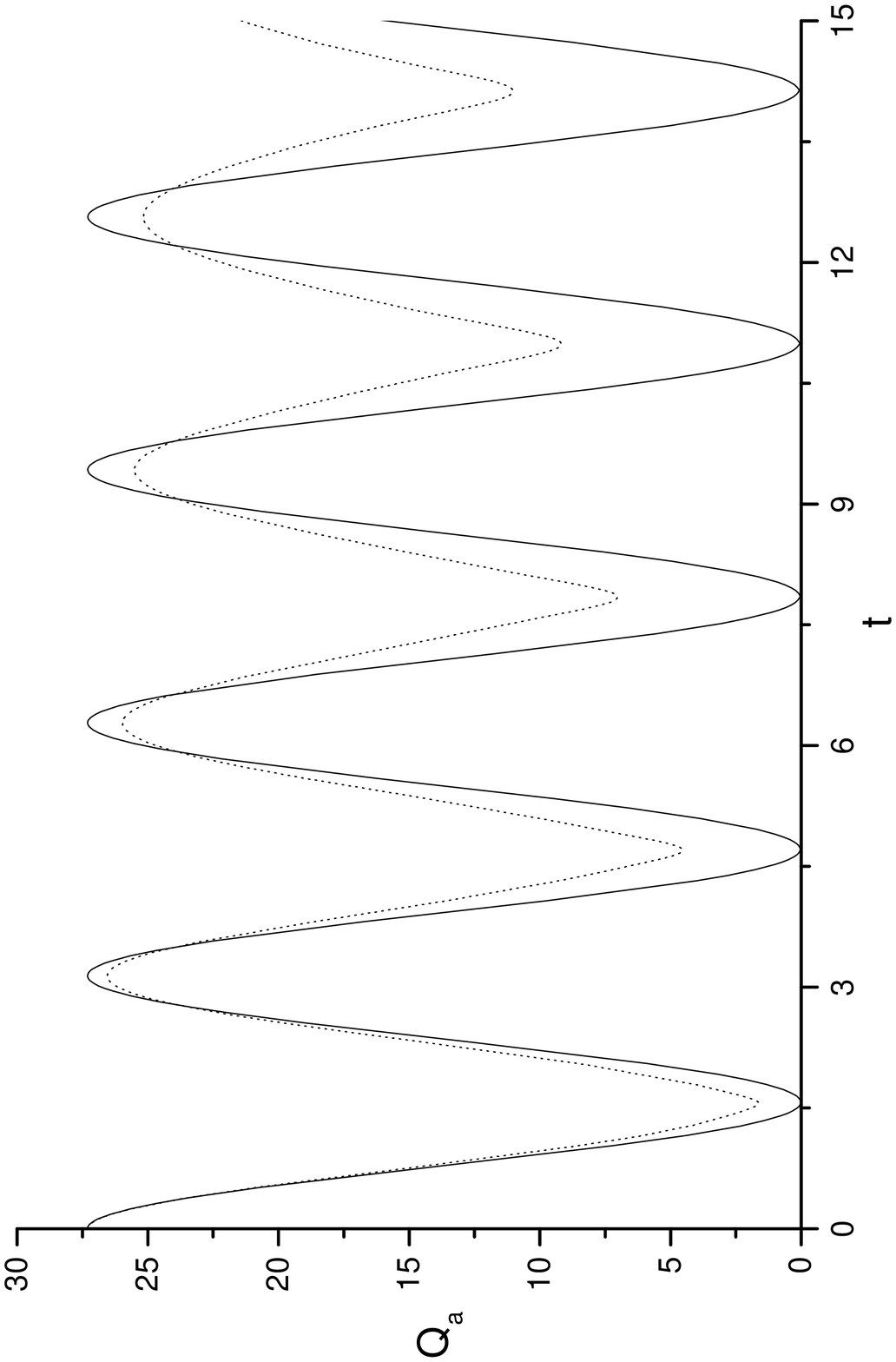}
\caption{The Mandel Q parameter $Q_a$ of optical field is plotted
as a function of time $t$ with $\Omega^{\prime}=1$ and $r=2$ for
two different values of the parameter $\gamma$, (Solid line)
$\gamma=\infty$, (Dot line) $\gamma=10^2$. \label{Fig1}}
\end{figure}

\begin{figure}
\centering
\includegraphics{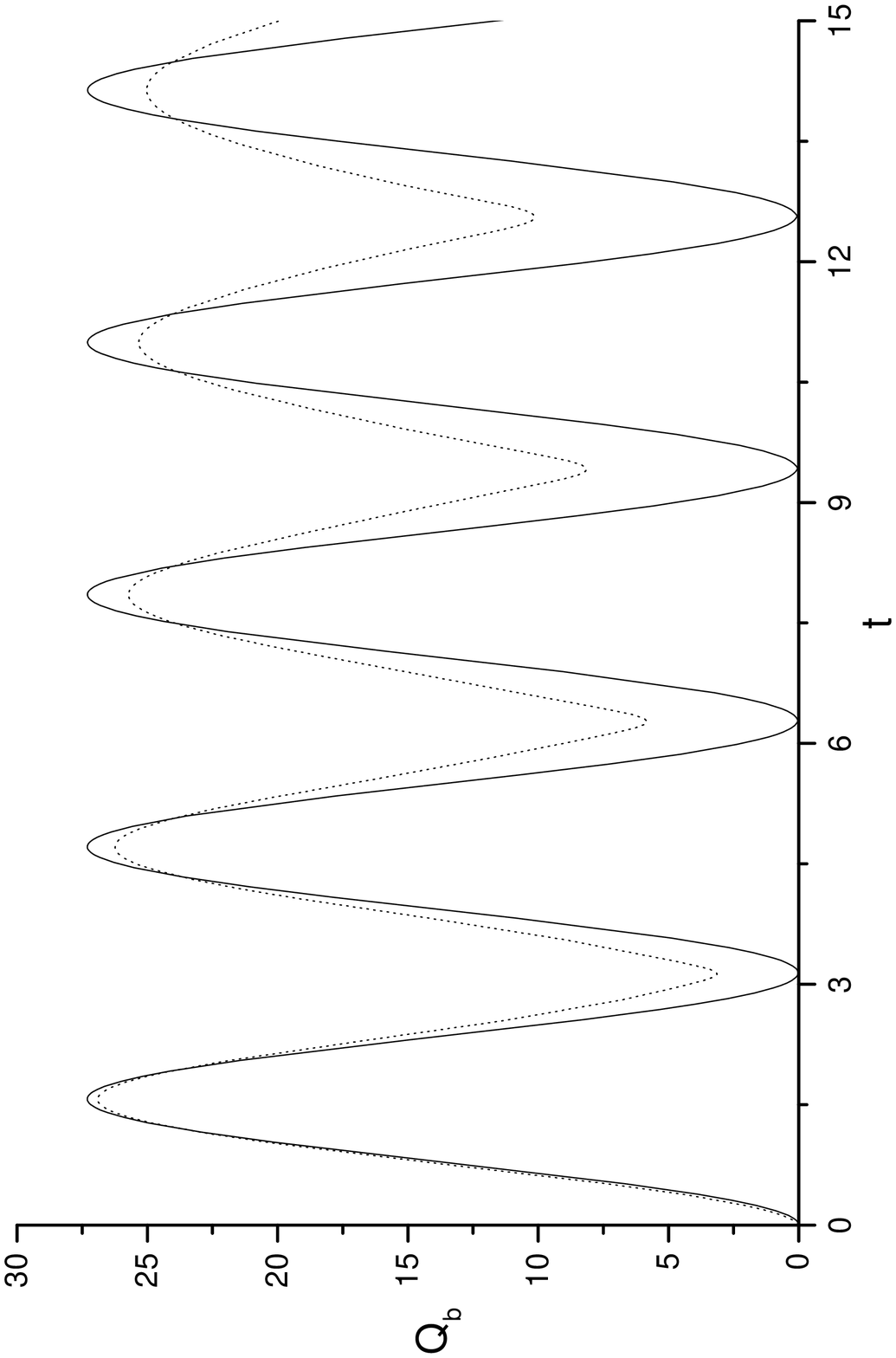}
\caption{The Mandel Q parameter $Q_b$ of atom laser field is
plotted as a function of time $t$ with $\Omega^{\prime}=1$ and
$r=2$ for two different values of the parameter $\gamma$, (Solid
line) $\gamma=\infty$, (Dot line) $\gamma=10^2$. \label{Fig2}}
\end{figure}

\begin{figure}
\centering
\includegraphics{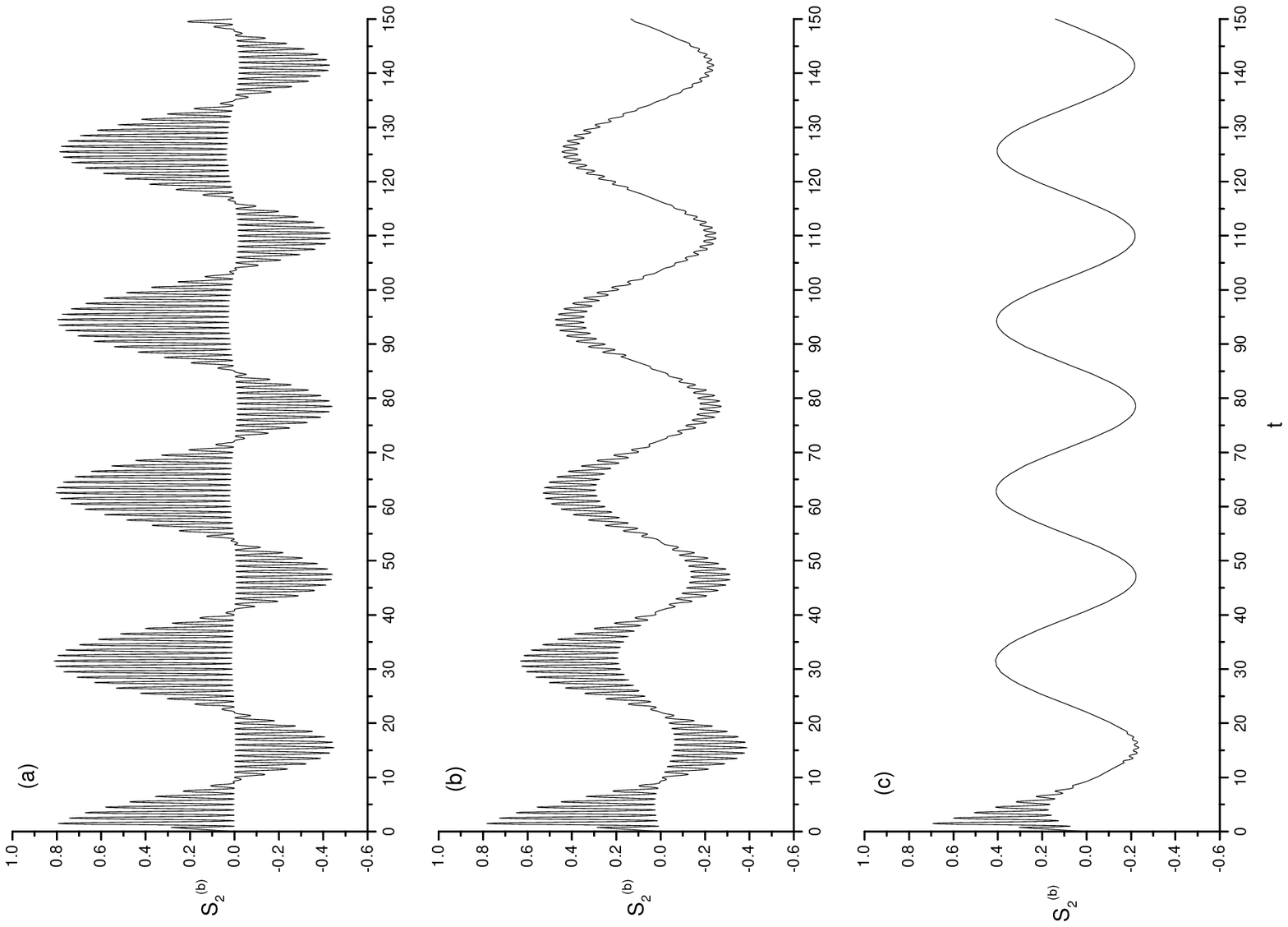}
\caption{The quadrature squeezing coefficient $S^{(b)}_2$ of the
atom laser field is plotted as a function of time $t$ with
$\omega=0.1$, $\phi=0$, $\theta=0$, $\Omega^{\prime}=\pi$ and
$r=0.3$ for three different values of the parameter $\gamma$, (a)
$\gamma=\infty$, (b) $\gamma=10^3$, (c) $\gamma=10^2$.
\label{Fig3}}
\end{figure}

\begin{figure}
\centering
\includegraphics{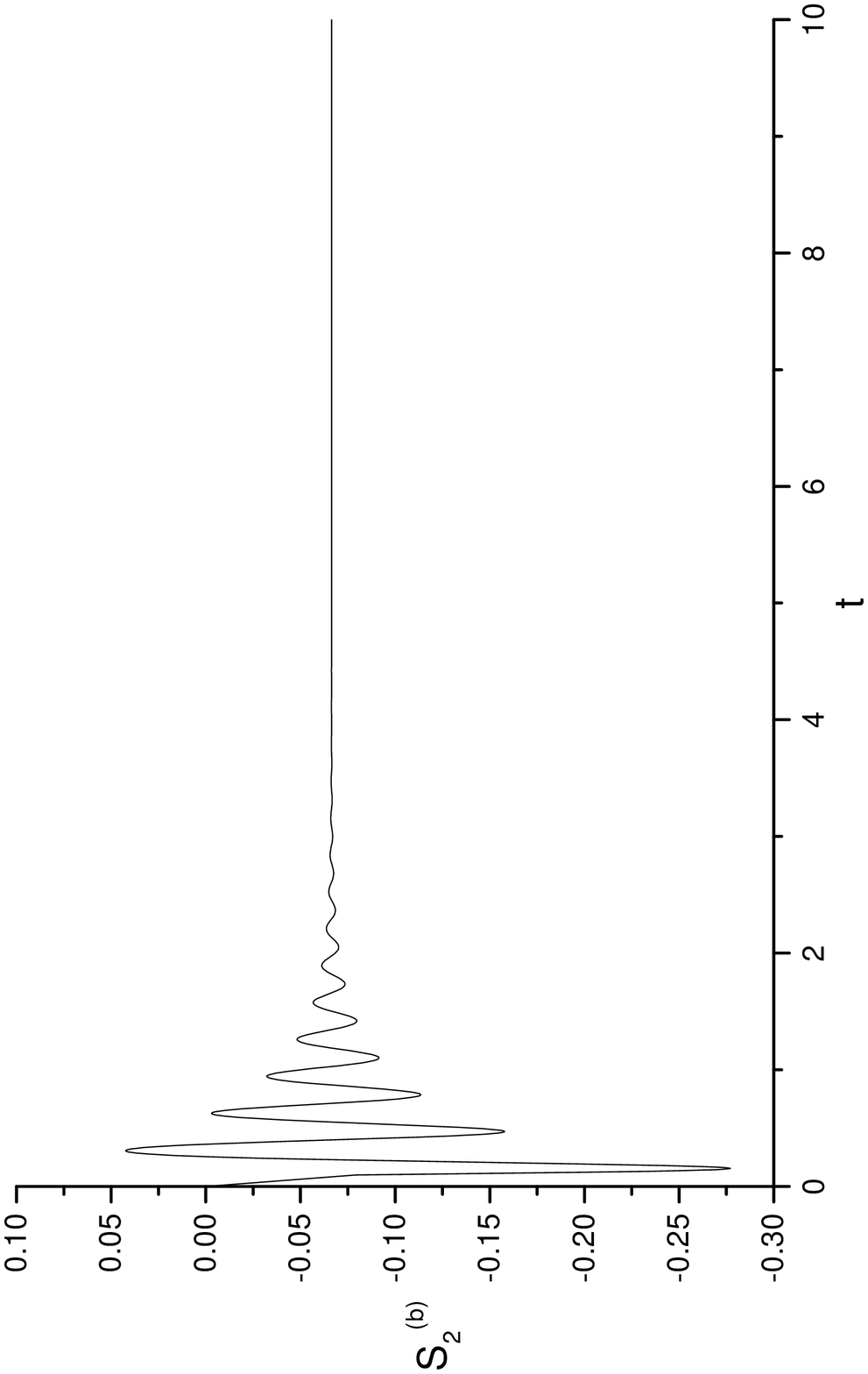}
\caption{The quadrature squeezing coefficient $S^{(b)}_2$ of the
atom laser field is plotted as a function of time $t$ with
$\omega=10$, $\phi=0$, $\theta=0$, $\Omega^{\prime}=10$, $r=0.3$
and $\gamma=10^2$. \label{Fig4}}
\end{figure}

\begin{figure}
\centering
\includegraphics{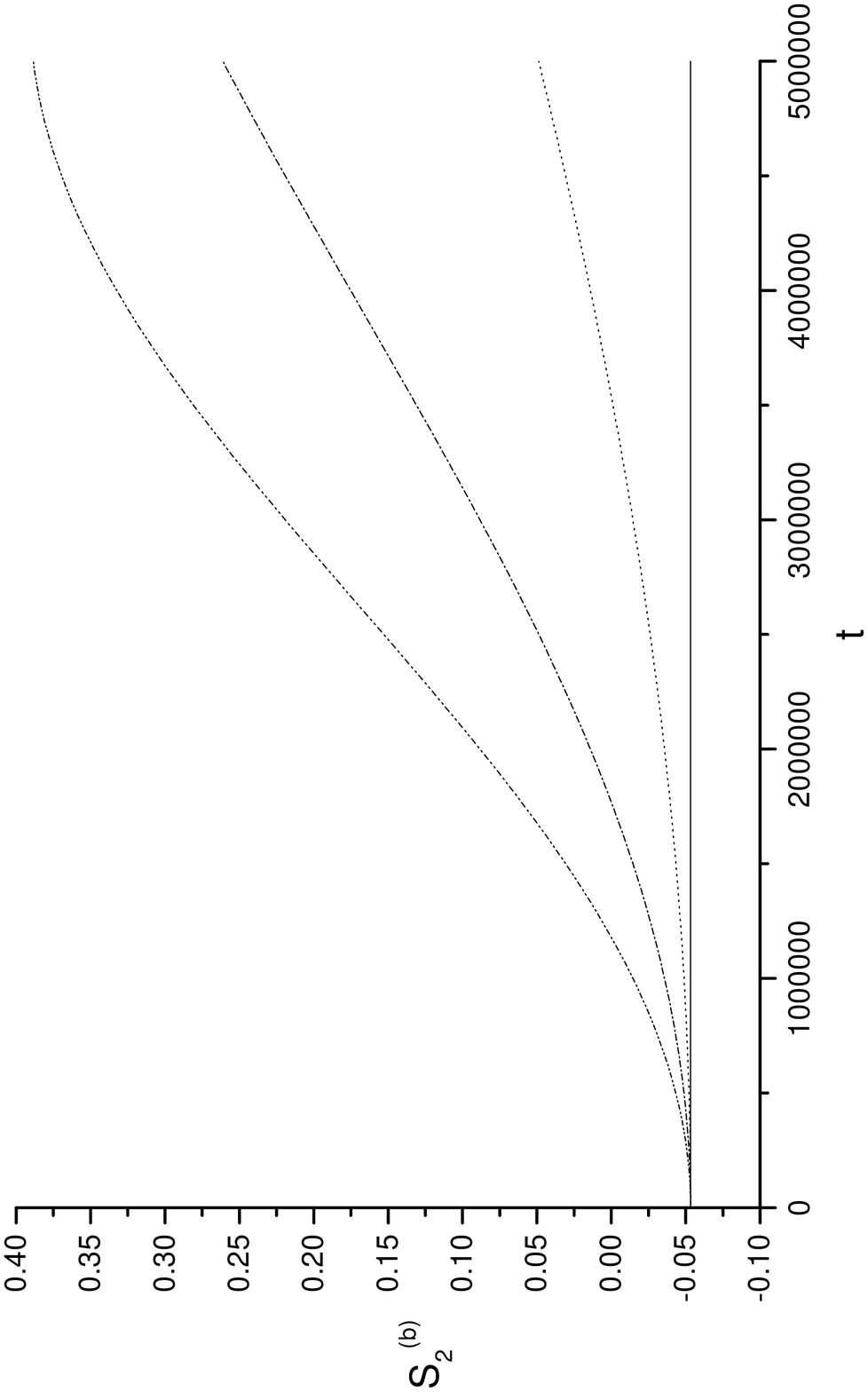}
\caption{The quadrature squeezing coefficient $S^{(b)}_2$ of the
atom laser field is plotted as a function of time $t$ with
$\omega=10$, $\phi=0$, $\theta=0$, $r=0.4$ and $\gamma=10^2$ for
four different values of $\Omega^{\prime}$: (Solid line)
$\Omega^{\prime}=10$, (Dot line) $\Omega^{\prime}=10+10^{-7}$,
(Dash Dot line) $\Omega^{\prime}=10+2\times10^{-7}$, (Dash Dot Dot
line) $\Omega^{\prime}=10+3\times10^{-7}$. \label{Fig5}}
\end{figure}

\end{document}